\documentclass{article}
\usepackage{geometry}                
\usepackage{amsmath,amssymb,amsthm,times,url}
\usepackage[sort&compress]{natbib}
\usepackage{graphicx}
\usepackage{epstopdf}
\def\today{\number\day\ \ifcase\month\or
 January\or February\or March\or April\or May\or June\or
 July\or August\or September\or October\or November\or December\fi
 \space \number\year}
\def\d{{\rm d}}
\def\e{{\rm e}}
\def\i{\ifmmode{\rm i}\else\char"10\fi}

\def\GG{\Omega}
\def\ph{\varphi}
\def\bfk{\boldsymbol{\xi}}

\def\dww{dispersive water wave}
\def\mbq{modified Boussinesq}
\def\bq{Boussinesq}

\def\p{Painlev\'e}
\def\sch{Schr\"odinger}
\def\bk{B\"ack\-lund}

\def\bts{B\"ack\-lund transformations}

\def\peq{\p\ equation}
\def\peqs{\p\ equations}

\def\PI{\mbox{\rm P$_{\rm I}$}}
\def\PII{\mbox{\rm P$_{\rm II}$}}
\def\PIII{\mbox{\rm P$_{\rm III}$}}
\def\PIV{\mbox{\rm P$_{\rm IV}$}}
\def\PV{\mbox{\rm P$_{\rm V}$}}
\def\PVI{\mbox{\rm P$_{\rm VI}$}}

\def\odes{ordinary differential equations}

\def\pdes{partial differential equations}
\newcommand{\deriv}[3][]{\frac{\d^{#1}{#2}}{{\d{#3}}^{#1}}}

\def\a{\alpha}
\def\b{\beta}

\def\la{\lambda}
\def\k{\kappa}
\def\ph{\varphi}

\def\th{\vartheta}
\def\VY{Yablonskii--Vorob'ev polynomials}
\def\Ok{Okamoto polynomials}
\def\ifrac#1#2{{#1}/{#2}}

\newcommand{\Integer}{\mathbb{Z}}

\def\Pf#1{\begin{proof}
#1\end{proof}}

\newcommand{\WhitD}[1]{D_{#1}}

\def\part#1#2%{\boldsymbol{\la}(#1,#2)}
{\boldsymbol{\la}({#1,#2})}

\def\hide#1{}
\newtheorem{theorem}{Theorem}
\newtheorem{definition}[theorem]{Definition}

\numberwithin{equation}{section}
\numberwithin{table}{section}
\numberwithin{figure}{section}
\numberwithin{theorem}{section}

\begin{document}

\title{Special Polynomials and Exact Solutions of the Dispersive Water Wave and Modified Boussinesq Equations}

% Names of the authors for the title of the paper
\author{Peter A.~Clarkson$^\dag$ and Bryn W.M.~Thomas$^\ddag$,\\
Institute of Mathematics, Statistics and Actuarial
Science,\\ University of Kent, Canterbury, CT2 7NF, UK\\
\texttt{$^\dag$P.A.Clarkson@kent.ac.uk}, $^\ddag$\texttt{bwmt3@kent.ac.uk}}
\maketitle 
\thispagestyle{empty}

\begin{abstract}
Exact solutions of the \dww\ and \mbq\ equations are expressed in terms of special polynomials associated with rational solutions of the fourth \p\ equation, which arises as generalized scaling reductions of these equations. Generalized solutions that involve an infinite sequence of arbitrary constants are also derived which are analogues of generalized rational solutions for the Korteweg-de Vries, Boussinesq and nonlinear \sch\ equations.\end{abstract}

\thispagestyle{plain}

\section{Introduction}
In this paper we are concerned with special polynomials associated with exact solutions of the \dww\ (DWW) equation
\begin{equation}\label{eq:dww}
U_{tt} + 2U_tU_{xx}+4U_xU_{xt}+6U_x^2U_{xx}-U_{xxxx}=0,
\end{equation}
which is a soliton equation solvable by inverse scattering \cite{refKaup75a,refKaup75b}, sometimes known as ``Kaup's higher-order wave equation" (cf.\ \cite{refPelSLL}), and the \mbq\ equation
\begin{equation}\label{eq:mbq}
\tfrac13U_{tt} - 2U_tU_{xx}-6U_x^2U_{xx}+U _{xxxx}=0,
\end{equation}
which also is a soliton equation solvable by inverse scattering \cite{refQCN}. These equations may be written in the non-local form (by setting $U_x=u$)
\begin{align}\label{eq:dwwnl}
&u_{tt} - 2u_{xx}\partial_x^{-1}(u_t)+4uu_{xt}+6u_xu_{t}+2(u^3)_{xx}-u_{xxxx}=0,\\
&\tfrac13u_{tt} - 2u_tu_x+2u_{xx}\partial_x^{-1}(u_t)-2(u^3)_{xx}+u_{xxxx}=0,\label{eq:mbqnl}
\end{align}
where $(\partial_x^{-1}f)(x)=\int_x^\infty f(y)\,\d y$, respectively, which is the form in which they arise in physical applications.
%The exact solutions of (\ref{eq:dww}) and (\ref{eq:mbq}) yield rational solutions of (\ref{eq:dwwnl}) and (\ref{eq:mbqnl}).

The DWW equation (\ref{eq:dww}) can be derived from the classical Boussinesq system
%\begin{subequations}\label{eq:clbq}
\begin{align}\label{eq:clbq}
&\eta_t+v_x+\eta\eta_x=0,\qquad
v_t+(\eta v)_x+\eta_{xxx}=0,
\end{align}%\end{subequations}
which arise in the description of surface waves propagating in shallow water \cite{refBroer,refKaup75a,refKaup75b,refWhit}. Indeed Broer \cite{refBroer} called the system ``the oldest, simplest and most widely known set of equations ...\ which are the Boussinesq equations proper". 
Hirota and Satsuma \cite{refHirSat} showed that there is a ``Miura type'' transformation relating solutions of the \mbq\ equation (\ref{eq:mbq}) to solutions of the Boussinesq equation
\begin{equation}\label{eq:bq}
u_{tt}+u_{xx}+(u^2)_{xx}+u_{xxxx}=0.
\end{equation}

There has been considerable interest in \pdes\ solvable by inverse
scattering, the \textit{soliton equations}, since the discovery in 1967
by Gardner, Greene, Kruskal and Miura \cite{refGGKM} of the method for
solving the initial value problem for the Korteweg-de Vries (KdV) equation
\begin{equation} u_t + 6uu_x + u_{xxx} =0.\label{eq:kdv} \end{equation} 
Clarkson and Ludlow \cite{refCL94} show that the generalized Boussinesq equation
\begin{equation}\label{eq:gbq}
U_{tt} + pU_tU_{xx}+qU_xU_{xt}+rU_x^2U_{xx}+U_{xxxx}=0,
\end{equation}
with $p$, $q$ and $r$ constants, satisfies the necessary conditions of the \p\ conjecture due to Ablowitz, Ramani and Segur \cite{refARSii,refARSiii} to be solvable by inverse scattering in two cases: (i), if $q=2p$ and $r=\tfrac32p^2$, when (\ref{eq:gbq}) is equivalent to the DWW equation (\ref{eq:dww}), and (ii), if $q=0$ and $r=-\tfrac12p^2$, when (\ref{eq:gbq}) is equivalent to the \mbq\ equation (\ref{eq:mbq}). 
%Hence it is thought that 

During the past thirty years or so there has been much interest in rational solutions of the soliton equations. %For some soliton equations solitons are given by rational solutions, e.g.\ for the Benjamin-Ono equation \cite{refMatsuno79,refSatIsh79} 
%and the Kadomtsev-Petviashvili equation \cite{refMZBIM,refSatAb79}.
Further applications of rational solutions to soliton equations include
the description of explode-decay waves \cite{refHN85} and vortex solutions of
the complex sine-Gordon equation \cite{refBP98,refOB05}.
The idea of studying the motion of poles of solutions of the KdV equation (\ref{eq:kdv}) is attributed to Kruskal \cite{refKruskal}. Airault, McKean and Moser \cite{refAMM77} studied the motion of the poles of rational solutions of the KdV equation (\ref{eq:kdv}) and the \bq\ equation (\ref{eq:bq}).
%\begin{equation}\label{eq:bq}
%u_{tt}+(u^2)_{xx}\pm u_{xxxx}=0,%\qquad\sigma^2=\pm1.
%\end{equation}
Further they related the motion to an integrable many-body problem, the Calogero-Moser system with constraints; see also \cite{refAdMoser,refCC77}. Ablowitz and Satsuma \cite{refASat} derived some rational solutions of the KdV equation (\ref{eq:kdv}) and the \bq\ equation (\ref{eq:bq}) by finding a long-wave limit of the known $N$-soliton solutions of these equations. Studies of rational solutions of other soliton equations include for the \bq\ equation \cite{refPACbqrats,refGPS,refPel98} and for the nonlinear \sch\ (NLS) equation \cite{refPAC06nls,refHone97,refHN85} %(\ref{eq:nls}) 
\begin{equation}\label{eq:nls}
\mbox{i} u_t + u_{xx} \pm2|u|^2u=0.
\end{equation}

Ablowitz and Segur \cite{refAS77} demonstrated a close relationship
between completely integrable \pdes\ solvable by inverse scattering and the
\peqs. For example the second \peq\ (\PII), 
\begin{equation}
\label{eq:PII} w'' = 2w^3 + z w + \a,
\end{equation} 
where $'\equiv\d/\d z$ and $\a$ is an arbitrary constant, 
arises as a scaling reduction of the KdV
equation (\ref{eq:kdv}), see \cite{refAS77}, and the fourth \peq\ (\PIV),
\begin{equation}
\label{eq:PIV} w'' = \frac{\left(w'\right)^{\!2}}{2w} +
\frac{3}{2}w^3 + 4z w^2 + 2(z^2 - \a)w + \frac{\b}{w},
\end{equation}
where $\a$ and $\b$ are arbitrary constants, arises as scaling reductions of the \bq\ equation (\ref{eq:bq}) and the NLS equation (\ref{eq:nls}). Consequently special solutions of these equations can be expressed in terms of solutions of \PII\ and \PIV.

The six \peqs\ (\PI--\PVI) are nonlinear \odes, the solutions of which are called the \emph{\p\ transcendents}, were discovered about a hundred years ago by
\p, Gambier and their colleagues whilst studying which second-order ordinary differential
equations have the property that the solutions have no movable branch points, i.e.\
the locations of multi-valued singularities of any of the solutions are
independent of the particular solution chosen and so are dependent only
on the equation; this is now known as the \emph{\p\ property}. \p,
Gambier\ \emph{et al.}\ showed that there were fifty canonical equations with this property, forty four are either integrable in terms of previously known
functions, such as elliptic functions or are equivalent to linear
equations, or are reducible to one of six new nonlinear ordinary differential
equations, which define new transcendental functions (cf.\ \cite{refInce}). 
The \p\ equations can be thought of as nonlinear analogues of the
classical special functions (cf.\
\cite{refPAC05review,refFIKN,refIKSY,refUmemura98}). Indeed
Iwasaki, Kimura, Shimomura and Yoshida \cite{refIKSY} characterize the
\peqs\ as ``the most important nonlinear \odes" and state that ``many
specialists believe that during the twenty-first century the \p\
functions will become new members of the community of special functions".
Further Umemura \cite{refUmemura98} states that ``Kazuo Okamoto and his circle
predict that in the 21st century a new chapter on \peqs\ will be
added to Whittaker and Watson's book". \hide{It is well known
that \PII--\PVI\ have rational solutions, algebraic solutions and
solutions expressed in terms of the classical special functions, though these solutions do not depend on two arbitrary constants and so are special solutions, sometimes known as ``\emph{classical solutions}" 
(see, e.g.\
\cite{refAirault,refBCH95,refPAC05review,refDM,refFA82,refGLS02,% 
refKLM,refMazz01a,refMazz01b,refMCB97,refMurata85,refMurata95,%
refOkamotoiii,refOkamotoi,refOkamotoii,refOkamotoiv} and the
references therein). }%

Vorob'ev \cite{refVor} and Yablonskii \cite{refYab59} expressed the rational solutions of \PII\ (\ref{eq:PII}) in terms of certain special polynomials, which are now known as the
\emph{\VY}. Okamoto \cite{refOkamotoiii} derived analogous special polynomials, which are now known as the \emph{\Ok}, related to some of the
rational solutions of \PIV\ (\ref{eq:PIV}). Subsequently Okamoto's results were generalized by Noumi and Yamada \cite{refNY99i} who showed that all rational solutions of \PIV\ can be expressed in terms of logarithmic derivatives of two sets of special polynomials, called the \emph{generalized Hermite polynomials} and the \emph{generalized Okamoto polynomials}\  (see \S\ref{sec:PIV} below). Clarkson and Mansfield \cite{refCM03} investigated the locations of the roots of the \VY\ in the complex plane and showed that these roots have a very regular, approximately triangular structure. The structure of the (complex) roots of the generalized Hermite and generalized Okamoto polynomials is described in \cite{refPAC03piv}, which respectively have an approximate rectangular structure and a combination of approximate rectangular and triangular structures. The term ``approximate" is used since the patterns are not exact triangles and rectangles as the roots lie on arcs rather than straight lines.

In this paper our interest is in exact solutions and associated polynomials of the special case of the DWW equation (\ref{eq:dww}) and the \mbq\ equation (\ref{eq:mbq}), both of which have generalized scaling reductions to \PIV\ (\ref{eq:PIV}). Consequently solutions of (\ref{eq:dww}) and (\ref{eq:mbq}) can be obtained in terms the generalized Hermite and generalized Okamoto polynomials. 
Further some of these solutions whose derivatives decay as $x\to\pm\infty$, are generalized to give more general solutions of the DWW equation (\ref{eq:dww}) and the \mbq\ equation (\ref{eq:mbq}).
These solutions are analogues of the rational solutions of the KdV equation %(\ref{eq:kdv}) 
\cite{refASat,refAdMoser,refAMM77,refCC77}, the \bq\ equation %(\ref{eq:bq}) 
\cite{refPACbqrats,refGPS,refPel98} and the NLS equation %(\ref{eq:nls}) 
\cite{refPAC06nls,refHone97}; see also \cite{refPAC06cmft}.
This paper is organized as follows. In \S\ref{sec:PIV} we review the special polynomials associated with rational solutions of  \PIV\ (\ref{eq:PIV}). In \S\S\ref{sec:dww} and \ref{sec:mbq} we use the special polynomials discussed in \S\ref{sec:PIV} to derive special polynomials and associated solutions of the DWW equation (\ref{eq:dww}) and the \mbq\ equation (\ref{eq:mbq}), respectively. We also derive generalized solutions which involve an infinite number of arbitrary constants.  All exact solutions of equations (\ref{eq:dww}) and (\ref{eq:mbq}), which are rational solutions of equations (\ref{eq:dwwnl}) and (\ref{eq:mbqnl}), which are described here are expressed as Wronskians of polynomials. Finally in \S\ref{sec:Dis} we discuss our results.

\section{\label{sec:PIV}Rational solutions of \PIV} 
%\subsection{Rational solutions of \PIV}
Simple rational solutions of \PIV\ (\ref{eq:PIV}) are
\begin{equation}\label{eq:pivsimprats}
w_1(z;\pm2,-2)=\pm\ifrac{1}{z},\quad
 w_2(z;0,-2)=-2z,\quad
w_3(z;0,-\tfrac29)=-\tfrac{2}{3}z.
\end{equation} It is known that there are three sets of
rational solutions of \PIV, which have the solutions
(\ref{eq:pivsimprats}) as the simplest members. These sets are known
as the ``$-1/z$ hierarchy", the ``$-2z$ hierarchy" and the ``$-\tfrac23z$
hierarchy", respectively (cf.\ \cite{refBCH95}). The ``$-1/z$
hierarchy" and the ``$-2z$ hierarchy" form the set of rational solutions
of \PIV\ (\ref{eq:PIV}) with parameters given by (\ref{eq:pivrats1}) and the
``$-\tfrac23z$ hierarchy" forms the set with parameters given by
(\ref{eq:pivrats2}). The rational solutions of \PIV\ (\ref{eq:PIV}) with parameters 
given by (\ref{eq:pivrats1}) lie at the vertexes of the ``Weyl
chambers" and those with parameters given by (\ref{eq:pivrats2}) lie
at the centres of the ``Weyl chamber" \cite{refUW97}.
%Rational solutions of \PIV\ (\ref{eq:PIV}) are summarized in the following theorem.

\begin{theorem}{\label{eq:pivrats}
\PIV\ (\ref{eq:PIV}) has rational solutions if and only if the parameters
$\a$ and $\b$ are given by either
\begin{subequations}\label{eq:pivrats12}\begin{equation} \label{eq:pivrats1}
\a=m,\qquad\b=-2(2n-m+1)^2,
\end{equation} or \begin{equation} 
\label{eq:pivrats2}
\a=m,\qquad\b=-2(2n-m+\tfrac13)^2,
\end{equation}\end{subequations} with $m,n\in\Integer$. 
For each given $m$ and $n$ there exists only one rational solution of \PIV\ with parameters  given by (\ref{eq:pivrats12}).
%Further the rational solutions forthese parameters are unique. 
}\end{theorem}

\Pf{See Lukashevich \cite{refLuk67a}, Gromak \cite{refGromak87}
and Murata \cite{refMurata85}; also Bassom, Clarkson and Hicks
\cite{refBCH95}, Gromak, Laine and Shimomura \cite[\S26]{refGLS02},
Umemura and Watanabe \cite{refUW97}.
%also \cite{refBCH95,refGLS02,refUW97}.
}

In a comprehensive study of properties of solutions of \PIV\ (\ref{eq:PIV}), Okamoto
\cite{refOkamotoiii} introduced two sets of polynomials associated with
rational solutions of \PIV\ (\ref{eq:PIV}), analogous to the {Yablonskii--Vorob'ev polynomials} associated with
rational solutions of \PII. Noumi and Yamada \cite{refNY99i} generalized Okamoto's results and introduced the \emph{generalized Hermite polynomials}, which are defined in Definition \ref{genherm}, and the \emph{generalized \Ok}, which are defined in Definition \ref{genokpoly}; see also \cite{refPAC03piv,refPAC06cmft}. Kajiwara and Ohta \cite{refKO98} expressed the generalized Hermite and generalized \Ok\ in terms of Schur functions in the form of determinants, which is how they are defined below. 

\begin{definition}\label{genherm}{\rm The \textit{generalized Hermite polynomial}\ $H_{m,n}(z)$ is defined by
\begin{align}\label{eq:eqHmn}H_{m,n}(z)&=\mathcal{W}\left(H_{m}(z), H_{m+1}(z), \ldots,H_{m+n-1}(z)\right)
%\equiv\mathcal{W}\left(\big\{H_{m+j}(z)\big\}_{j=0}^{n-1}\right)
, \end{align} 
for $m,n\geq1$, with $H_{m,0}(z)=H_{0,n}(z)=1$ and $H_{k}(z)$ the $k$th Hermite polynomial.
%and has degree $\mbox{deg}(H_{m,n}(z))=mn$.
}\end{definition}

The polynomials, $H_{m,n}(z)$, defined by (\ref{eq:eqHmn}) are called the
{generalized Hermite polynomials} since $H_{m,1}(z)=H_{m}(z)$ and
$H_{1,m}(z)=\i^{-m}H_{m}(\i z)$, where $H_{m}(z)$ is the standard Hermite
polynomial defined by
\begin{equation*}H_{m}(z)=(-1)^m\exp(z^2)\frac{\d^m}{\d z^m}
\left\{\exp(-z^2)\right\}\end{equation*} or alternatively through the
generating function
\begin{equation}\label{genfun}\sum_{m=0}^\infty\frac{H_{m}(z)\,\xi^m}{m!} 
=\exp(2\xi z-\xi^2).\end{equation} %(cf.\ \cite{refAbSt}). 

Examples of generalized Hermite polynomials and plots of the locations of their roots in the complex plane are given by Clarkson \cite{refPAC03piv}; see also \cite{refPAC06cmft,refPAC06nls,refPACbqrats}. The roots take the form of $m\times n$ ``rectangles", which are only approximate rectangles since the roots lie on arcs rather than straight lines. 
The generalized Hermite polynomial $H_{m,n}(z)$ can be expressed as the multiple integral
\begin{align*}H_{m,n}(z) = \frac{\pi^{m/2}\prod_{k=1}^mk!}{2^{m(m+2n-1)/2}}
\int_{-\infty}^{\infty}&\!\!\!\begin{array}{c}{\cdots}\\[-10pt]{}_{n}\end{array}\!\!
\int_{-\infty}^{\infty}
\prod_{i=1}^{n}\prod_{j=i+1}^{n}(x_i-x_j)^2%\\&\times
\prod_{k=1}^n(z-x_k)^m \exp\left(-x_k^2\right)
\d x_1\,\d x_2\ldots\d x_n,\end{align*}
which arises in random matrix theory \cite{refBH00,refFW01, refKanzieper}. 
The generalized Hermite polynomials also arise in the theory of orthogonal polynomials \cite{refCF06}.

\begin{theorem}\label{genhermthm}{Suppose $H_{m,n}(z)$ is the generalized Hermite polynomial, then
\begin{subequations}\label{eq:p4ratsols12}\begin{align} \label{p4ratsols1}
w_{m,n}^{[1]} (z)
&=\deriv{}{z}\ln\left\{\frac{H_{m+1,n}(z)}{H_{m,n}(z)}\right\},\\
w_{m,n}^{[2]}(z)
&=\deriv{}{z}\ln\left\{\frac{H_{m,n}(z)}{H_{m,n+1}(z)}\right\},\\
w_{m,n}^{[3]}(z)&=
-2z+\deriv{}{z}\ln\left\{\frac{H_{m,n+1}(z)}{H_{m+1,n}(z)}\right\}, 
\end{align}\end{subequations} where $w_{m,n}^{[j]} =
w(z;\a^{[j]}_{m,n},\b^{[j]}_{m,n})$, $j=1,2,3$, are solutions of \PIV,
respectively for 
\begin{align*}
\a_{m,n}^{[1]}&=2m+n+1,&\qquad\b_{m,n}^{[1]}&=-2n^2,\\
\a_{m,n}^{[2]}&=-(m+2n+1),&\qquad\b_{m,n}^{[2]}&=-2m^2,\\ 
\a_{m,n}^{[3]}&=n-m,&\qquad\b_{m,n}^{[3]}&=-2(m+n+1)^2.
\end{align*}
}\end{theorem}

\Pf{See Theorem 4.4 in Noumi and Yamada
\cite{refNY99i}; also Theorem 3.1 in \cite{refPAC03piv}.}

The rational solutions of
\PIV\ defined by (\ref{eq:p4ratsols12}) include all solutions in the
``$-1/z$" and ``$-2z$" hierarchies, i.e.\ the set of rational solutions
of \PIV\ with parameters given by (\ref{eq:pivrats1}). In fact these rational solutions of \PIV\ (\ref{eq:PIV}) are special cases of the special function solutions which are expressible in terms of parabolic cylinder functions $\WhitD{\nu}(\xi)$.

\begin{definition}\label{genokpoly}{\rm The \textit{generalized Okamoto polynomial} $\GG_{m,n}(z)$ is defined by
\begin{subequations}\begin{align}
\GG_{m,n}(z) &=\mathcal{W}(H_1(z),H_4(z),\ldots,H_{3m-2}(z),H_2(z),H_5(z),\ldots,H_{3n-1}(z)),\\
%&\equiv\mathcal{W}\left(\big\{H_{3j+1}(z)\big\}_{j=0}^{m-1},\big\{H_{3k+2}(z)\big\}_{k=0}^{n-1}\right),
%\nonumber\\
\GG_{m,0}(z) &=\mathcal{W}(H_1(z),H_4(z),\ldots,H_{3m-2}(z))
%\equiv\mathcal{W}\left(\big\{H_{3j+1}(z)\big\}_{j=1}^{m-1}\right)
,\\
\GG_{0,n}(z) &=\mathcal{W}(H_2(z),H_5(z),\ldots,H_{3n-1}(z)),
%\equiv\mathcal{W}\left(\big\{H_{3k+2}(z)\big\}_{j=1}^{n-1}\right),
\end{align}\end{subequations}for $m,n\geq1$, 
with $\GG_{0,0}(z)=1$ and $H_{k}(z)$ the $k$th Hermite polynomial.
%and has degree $\mbox{deg}(\GG_{m,n}(z))=m^2+n^2-mn+n$.
}\end{definition}

The generalized Okamoto polynomial $\GG_{m,n}(z)$ defined here have been reindexed in comparison to the generalized Okamoto polynomial $Q_{m,n}(z)$ defined in \cite{refPAC03piv,refPAC06cmft} by setting $Q_{m,n}(z)=\GG_{m+n-1,n-1}(z)$ and $Q_{-m,-n}(z)=\GG_{n-1,m+n}(z)$, for $m,n\geq1$. The polynomials introduced by Okamoto \cite{refOkamotoiii} are given by $Q_{m}(z)=\GG_{m-1,0}(z)$ and $R_{m}(z)=\GG_{m,1}(z)$. Further the generalized Okamoto polynomial introduced by Noumi and Yamada \cite{refNY99i} is given by $\widetilde{Q}_{m,n}(z)=\GG_{m-1,n-1}(z)$.

Examples of generalized \Ok\ and plots of the locations of their roots in the complex plane are given by Clarkson \cite{refPAC03piv,refPAC06cmft}. The roots of the polynomial $Q_{m,n}(z)=\GG_{m+n-1,n-1}(z)$ with $m,n\geq1$ take the form of an $m\times n$ ``rectangle" with an ``equilateral triangle", which have either $\tfrac12m(m-1)$ or $\tfrac12n(n-1)$ roots, on each of its sides. The roots of the polynomial $Q_{-m,-n}(z)=\GG_{n-1,m+n}(z)$ with $m,n\geq1$ take the form of an $m\times n$ ``rectangle" with an ``equilateral triangle", which now have either $\tfrac12m(m+1)$ or $\tfrac12n(n+1)$ roots, on each of its sides. Again these are only approximate rectangles and equilateral triangles since the roots lie on arcs rather than straight lines. 

\begin{theorem}\label{genokpolythm}{Suppose $Q_{m,n}(z)$ is the generalized Okamoto polynomial,
then
\begin{subequations}\label{eq:eqgenokrats112}
\begin{align} \label{eqgenokrats11}
\widetilde{w}_{m,n}^{[1]} (z)&= -\tfrac23{z}
+\deriv{}{z}\ln\left\{\frac{Q_{m+1,n}(z)}{Q_{m,n}(z)}\right\}, \\ 
\label{eqgenokrats12} 
 \widetilde{w}_{m,n}^{[2]} (z)&= -\tfrac23{z}
+\deriv{}{z}\ln\left\{\frac{Q_{m,n}(z)}{Q_{m,n+1}(z)}\right\}, \\ 
\widetilde{w}_{m,n}^{[3]} (z)&= -\tfrac23{z}
+\deriv{}{z}\ln\left\{\frac{Q_{m,n+1}(z)}{Q_{m+1,n}(z)}\right\}, 
\end{align}\end{subequations}
where $\widetilde{w}_{m,n}^{[j]} =
w(z;\widetilde{\a}_{m,n}^{[j]},\widetilde{\b}_{m,n}^{[j]})$,
$j=1,2,3$, are solutions of \PIV, respectively for 
\begin{align*}\widetilde{\a}_{m,n}^{[1]}&=2m+n,&\qquad
\widetilde{\b}_{m,n}^{[1]}&=-2(n-\tfrac13)^2,\\ 
\widetilde{\a}_{m,n}^{[2]}&=-(m+2n),&\qquad
\widetilde{\b}_{m,n}^{[2]}&=-2(m-\tfrac13)^2,\\ 
\label{eqgenokrats15} \widetilde{\a}_{m,n}^{[3]}&=n-m,&\qquad
\widetilde{\b}_{m,n}^{[3]}&=-2(m+n+\tfrac13)^2.
\end{align*}
}\end{theorem}

\Pf{See Theorem 4.3 in Noumi and Yamada
\cite{refNY99i}; also Theorem 4.1 in \cite{refPAC03piv}. }

\section{\label{sec:dww}Exact solutions of the dispersive water wave equation} 
\subsection{Exact solutions from the scaling reduction}
%Clarkson and Ludlow  \cite{refCL94} show that 
The DWW equation (\ref{eq:dww}) has the generalized scaling reduction \cite{refCL94}
\begin{equation}\label{eq:dwwred}
U(x,t)=V(z)-\k x-\k^2t-\mu\ln t,\qquad z=\ifrac{(x+2\k t)}{\sqrt{4t}},\end{equation}
with $\k$ and $\mu$ arbitrary constants, which is a classical symmetry reduction \cite{refCL94}, and where $v(z)=V'(z)$ satisfies
\begin{equation}\label{eq19}
v''' = 6v^2v' - 12zvv' + (4z^2-8\mu)v'-8v^2+12zv +16\mu.
\end{equation}
Then letting $v(z)=w(z)+2z$ in (\ref{eq19}) and integrating yields \PIV\ (\ref{eq:PIV}) and so we can obtain exact solutions of the DWW equation (\ref{eq:dww}) from the rational solutions of \PIV\ given in \S\ref{sec:PIV}. However, it is possible to generate these solutions directly,
i.e.\ without having to consider the generalized Hermite and generalized
Okamoto polynomials, by extending the representations of these
polynomials in terms of the determinants given in Definitions \ref{genherm} and \ref{genokpoly}.

\begin{theorem}{\label{thm32}Consider the polynomials $\ph_n(x,t;\k)$ defined by
\begin{equation}\label{eq:phn}\sum_{n=0}^\infty
\frac{\ph_n(x,t;\k)\la^n}{n!}=\exp\left\{(x+2\k t)\la-t\la^2\right\},\end{equation} so
$\ph_n(x,t;\k) = t^{n/2}H_{n}\left(\ifrac{(x+2\k t)}{\sqrt{4t}}\right)$, 
with $H_n(z)$ the Hermite polynomial, and then let 
\begin{equation}\label{eq:Phimn}\Phi_{m,n}(x,t;\k) = %a_{m,n}
\mathcal{W}_x(\ph_m,\ph_{m+1},\ldots,\ph_{m+n-1}).\end{equation}
where $\mathcal{W}_x(\ph_m,\ph_{m+1},\ldots,\ph_{m+n-1})$ is
the Wronskian with respect to $x$. %, which is a polynomial in $x$ of
%degree $mn$ with coefficients that are polynomials in $t$ and has the
%symmetry property $\Phi_{m,n}(x,t)=\Phi_{n,m}(x,-t)$. 
%Hence it can be shown that 
Then the DWW equation (\ref{eq:dww}) has exact solutions in the form
\begin{subequations}\label{eq:dwwsols123}\begin{align}
U_{m,n}^{[1]}(x,t;\k)&=\ln\left\{\frac{\Phi_{m+1,n}(x,t;\k)}{\Phi_{m,n}(x,t;\k)}\right\}
+\frac{x^2}{4t}-(m+n+\tfrac12)\ln t,\\
U_{m,n}^{[2]} (x,t;\k)&=\ln\left\{\frac{\Phi_{m,n}(x,t;\k)}{\Phi_{m,n+1}(x,t;\k)}\right\}
+\frac{x^2}{4t}+(m+n+\tfrac12)\ln t,\\
U_{m,n}^{[3]}(x,t;\k)&=\ln\left\{\frac{\Phi_{m,n+1}(x,t;\k)}{\Phi_{m+1,n}(x,t;\k)}\right\}-\k(x+\k t),
\end{align}\end{subequations}
}\end{theorem}

\Pf{The polynomials $\ph_n(x,t;\k)$ defined by (\ref{eq:phn}) are obtained by $z=(x+\k t)/\sqrt{4t}$ and $\xi=\la\sqrt{t}$ in (\ref {genfun}). Then (\ref{eq:Phimn}) follows from the definition of $H_{m,n}(z)$ given by (\ref{eq:eqHmn}). Finally substitution of these expressions into (\ref{eq:dwwred}) yields the desired result.}

%We remark that the polynomials $\Phi_{m,n}(x,t)$ can be expressed in terms of the Schur polynomial $S_{\bY_{m,n}}(x,-t,0,0,\ldots)$ with partition $\bY_{m,n}=([m]^n)$.

\begin{theorem}{\label{thm33}Consider the polynomials ${\psi}_n(x,t;\k)$ defined by
\begin{equation}\label{eq:psn}\sum_{n=0}^\infty\frac{\psi_n(x,t;\k)\la^n}{n!}=
\exp\left\{(x+2\k t)\la+3t\la^2\right\},\end{equation} so
$\psi_n(x,t;\k)=(3t)^{n/2}\e^{-n\pi\i/2}\,H_n\left(\ifrac{\i (x+2\k t)}{\sqrt{3t}}\right)$,
with $H_n(z)$ the Hermite polynomial, and then let
\begin{equation}\label{eq:Psimn} \begin{split} \Psi_{m,n}(x,t;\k) &= 
\mathcal{W}_x\left(\psi_1,\psi_4,\ldots,\psi_{3m+3n-5},
\psi_2,\psi_5,\ldots,\psi_{3n-4}\right),\\ \Psi_{-m,-n}(x,t) &=
\mathcal{W}_x\left(\psi_1,\psi_4,\ldots,\psi_{3n-2},
\psi_2,\psi_5,\ldots,\psi_{3m+3n-1}\right),
\end{split}\end{equation}
for $m,n\geq1$, where $\mathcal{W}_x(\psi_1,\psi_{2},\ldots,\psi_{m})$ is
the Wronskian with respect to $x$. %which is a polynomial in $x$ of degree
%$m^2+n^2+mn-m-n$ with coefficients that are polynomials in $t$ and has
%the symmetry property $\Psi_{m,n}(x,t)=\Psi_{n,m}(x,-t)$. 
Then the DWW equation (\ref{eq:dww}) has exact solutions in the form
\begin{subequations}\label{eq:dwwsols456}\begin{align}%{l@{\quad}l}
U_{m,n}^{[1]} (x,t;\k)&=\ln\left\{\frac{\Psi_{m+1,n}(x,t)}{\Psi_{m,n}(x,t;\k)}\right\}
+\frac{x^2}{6t}-\tfrac13\k (x+\k t)-(2m+n)\ln t,\\
U_{m,n}^{[2]}(x,t;\k)&=
\ln\left\{\frac{\Psi_{m,n}(x,t;\k)}{\Psi_{m,n+1}(x,t;\k)}\right\}
+\frac{x^2}{6t}-\tfrac13\k(x+\k t)+(m+2n)\ln t,\\
U_{m,n}^{[3]}(x,t;\k)&=
\ln\left\{\frac{\Psi_{m,n+1}(x,t;\k)}{\Psi_{m+1,n}(x,t;\k)}\right\}
+\frac{x^2}{6t}-\tfrac13\k (x+\k t)-(m-n)\ln t.
\end{align}\end{subequations}
}\end{theorem}

\Pf{The proof is similar to that for Theorem \ref{thm32} above and so is left to the reader.}

\subsection{Generalized solutions}
Here we discuss possible generalizations of the rational solutions obtained above. 
Motivated by the structure of the generalized exact solutions of the KdV equation \cite{refAdMoser,refAMM77}, the \bq\ equation \cite{refPACbqrats,refGPS}, and the NLS equation \cite{refPAC06nls}, the idea is to replace the exponent of the exponentials (\ref{eq:phn}) and (\ref{eq:psn}) by the infinite series 
\begin{equation}\label{infseries}(x+2\k t)\la+bt\la^2+\sum_{j=3}^\infty \xi_j\la^j,\end{equation}
with $b=-1$ and $b=3$, respectively, where $\xi_j$, for $j\geq3$, are arbitrary parameters. These give generalizations of the polynomials (\ref{eq:Phimn}) and (\ref{eq:Psimn}) which in turn give generalizations of (\ref{eq:dwwsols123}) and (\ref{eq:dwwsols456}). However, when we substitute the generalizations of (\ref{eq:dwwsols123}) and (\ref{eq:dwwsols456}) into the DWW equation (\ref{eq:dww}), it can be shown that necessarily $\k=0$ and $\xi_j=0$ for $j\geq3$ except for the generalization of (\ref{eq:dwwsols123}c), and is described in the following theorem.

\begin{theorem}{Consider the polynomials $\theta_n(x,t;\bfk)$ defined by
\begin{equation}\sum_{n=0}^\infty
\frac{\th_n(x,t;\bfk)\la^n}{n!}=\exp\left(x\la-t\la^2+\sum_{j=3}^\infty
\xi_j\la^j\right),\end{equation} where $\bfk=(\xi_3,\xi_4,\ldots)$,
with $\xi_j$ arbitrary constants and then let %$\Phi_{m,n}(x,t;\bfk_{m+n-1})$ by
\begin{equation}\label{eq:Phimnk}
\Theta_{m,n}(x,t;\bfk) = %a_{m,n}
\mathcal{W}_x(\th_m,\th_{m+1},\ldots,\th_{m+n-1}).\end{equation}
where $\mathcal{W}_x(\theta_m,\theta_{m+1},\ldots,
\theta_{m+n-1})$ is the Wronskian with respect to $x$. %which is a polynomial in $x$ of degree
%$mn$ with coefficients that are polynomials in $t$ and has the symmetry
%property $\Theta_{m,n}(x,t;\bfk_{m+n-1})= \Theta_{n,m}(x,-t;\bfk_{m+n-1})$. 
Then %it can be shown that 
the DWW equation (\ref{eq:dww}) has exact solutions in the form
\begin{equation}\label{sols:genrats}
U_{m,n}^{[3]}(x,t;\bfk)=\ln
\left\{\ifrac{\Theta_{m,n+1}(x,t;\bfk)}
{\Theta_{m+1,n} (x,t;\bfk)}\right\},\end{equation} }\end{theorem}

The polynomials (\ref{eq:Phimnk}) and associated solutions (\ref{sols:genrats}) are analogous to the polynomials and associated rational solutions of the KdV equation
(\ref{eq:kdv}) derived by Airault, McKean and Moser \cite{refAMM77} and
Adler and Moser \cite{refAdMoser}; see also \cite{refASat,refCC77}.
Further we conclude that generalized solutions, i.e.\ solutions which depend on an infinite number of arbitrary parameters, of the DWW equation (\ref{eq:dww})
exist only if the derivative of the solution, which is a decaying rational solution of (\ref{eq:dwwnl}), obtained through the scaling reduction to a \peq\ decays as $|x|\to\infty$. This is analogous to the situation for generalized rational solutions of the KdV equation \cite{refAdMoser,refAMM77,refCC77}, the \bq\ equation \cite{refPACbqrats,refGPS,refPel98}, and the NLS equation \cite{refHone97}; see also \cite{refPAC06cmft}.

\section{\label{sec:mbq}Exact solutions of the modified Boussinesq equation} 
\subsection{Exact solutions from the scaling reduction}
The \mbq\ equation (\ref{eq:mbq}) has the generalized scaling reduction \cite{refPAC89mbq}
\begin{equation}\label{eq30}
U(x,t)=V(z)+\k xt-\mu\ln t,\qquad z=\ifrac{(x+3\k t^2)}{\sqrt{4t}},\end{equation}
with $\k$ and $\mu$ arbitrary constants and where $v(z)=V'(z)$ satisfies
\begin{equation}\label{eq31}
v''' = 6v^2v' +4zvv' - (\tfrac43z^2+8\mu)v'-4zv+\tfrac{16}3\mu.
\end{equation}
In the case when $\k\not=0$, the generalized scaling reduction (\ref{eq30}) is a nonclassical symmetry reduction that is derived either using the nonclassical method \cite{refBCi} or the direct method \cite{refCK} --- see \cite{refPAC89mbq} for details. Letting $V(z)=w(z)+\tfrac23z$ in (\ref{eq31}) and integrating yields \PIV\ (\ref{eq:PIV}) and so we can obtain exact solutions of the \mbq\ equation (\ref{eq:mbq}) 
from the rational solutions of \PIV\ given in \S\ref{sec:PIV}. As for the DWW equation (\ref{eq:dww}), it is possible to generate these solutions directly,
i.e.\ without having to consider the generalized Hermite and generalized
Okamoto polynomials, by extending the representations of these
polynomials in terms of the determinants given in Definitions \ref{genherm} and \ref{genokpoly}.

\begin{theorem}{\label{thm42}Consider the polynomials $\ph_n(x,t;\k)$ defined by
\begin{equation}\label{eq:phn2}\sum_{n=0}^\infty
\frac{\ph_n(x,t;\k)\la^n}{n!}=\exp\left\{(x+3\k t^2)\la-t\la^2\right\},\end{equation} so
$\ph_n(x,t;\k) = t^{n/2}H_{n}\left(\ifrac{(x+3\k t^2)}{\sqrt{4t}}\right)$, 
with $H_n(z)$ the Hermite polynomial, and then let 
\begin{equation}\label{eq:Phimn2}\Phi_{m,n}(x,t;\k) = %a_{m,n}
\mathcal{W}_x(\ph_m,\ph_{m+1},\ldots,\ph_{m+n-1}).\end{equation}
where $\mathcal{W}_x(\ph_m,\ph_{m+1},\ldots,\ph_{m+n-1})$ is
the Wronskian with respect to $x$. Then the \mbq\ equation (\ref{eq:mbq}) has exact solutions in the form
\begin{subequations}\label{eq:mbqsols123}\begin{align}
U_{m,n}^{[1]}(x,t;\k)&=\ln\left\{\frac{\Phi_{m+1,n}(x,t;\k)}{\Phi_{m,n}(x,t;\k)}\right\}
+\frac{x^2}{12t}+\tfrac32\k xt+\tfrac34\k^2t^3+(m+\tfrac12)\ln t,\\
U_{m,n}^{[2]} (x,t;\k)&=\ln\left\{\frac{\Phi_{m,n}(x,t;\k)}{\Phi_{m,n+1}(x,t;\k)}\right\}
+\frac{x^2}{12t}+\tfrac32\k xt+\tfrac34\k^2t^3-(n+\tfrac12)\ln t,\\
U_{m,n}^{[3]}(x,t;\k)&=\ln\left\{\frac{\Phi_{m,n+1}(x,t;\k)}{\Phi_{m+1,n}(x,t;\k)}\right\}
-\frac{x^2}{6t}-\tfrac32\k^2t^3-(m-n)\ln t,
\end{align}\end{subequations}
}\end{theorem}

\begin{theorem}{\label{thm43}Consider the polynomials ${\psi}_n(x,t;\k)$ defined by
\begin{equation}\label{eq:psn2}\sum_{n=0}^\infty\frac{\psi_n(x,t;\k)\la^n}{n!}=
\exp\left\{(x+3\k t^2)\la+3t\la^2\right\},\end{equation} so
$\psi_n(x,t;\k)=(3t)^{n/2}\e^{-n\pi\i/2}\,H_n\left(\ifrac{\i(x+3\k t^2)}{\sqrt{3t}}\right)$,
with $H_n(z)$ the Hermite polynomial, and then let
\begin{equation}\label{eq:Psimn2} \begin{split} \Psi_{m,n}(x,t;\k) &= 
\mathcal{W}_x\left(\psi_1,\psi_4,\ldots,\psi_{3m+3n-5},
\psi_2,\psi_5,\ldots,\psi_{3n-4}\right),\\ \Psi_{-m,-n}(x,t;\k) &=
\mathcal{W}_x\left(\psi_1,\psi_4,\ldots,\psi_{3n-2},
\psi_2,\psi_5,\ldots,\psi_{3m+3n-1}\right),
\end{split}\end{equation}
for $m,n\geq1$, where $\mathcal{W}_x(\psi_1,\psi_{2},\ldots,\psi_{m})$ is
the Wronskian with respect to $x$.
Then the \mbq\ equation (\ref{eq:mbq}) has exact solutions in the form
\begin{subequations}\label{eq:mbqsols456}\begin{align}%{l@{\quad}l}
U_{m,n}^{[1]} (x,t;\k)&=\ln\left\{\ifrac{\Psi_{m+1,n}(x,t;\k)}{\Psi_{m,n}(x,t;\k)}\right\}+\k xt,\\
U_{m,n}^{[2]}(x,t;\k)&=\ln\left\{\ifrac{\Psi_{m,n}(x,t;\k)}{\Psi_{m,n+1}(x,t;\k)}\right\}+\k xt,\\
U_{m,n}^{[3]}(x,t;\k)&=\ln\left\{\ifrac{\Psi_{m,n+1}(x,t;\k)}{\Psi_{m+1,n}(x,t;\k)}\right\}+\k xt.
\end{align}\end{subequations}
}\end{theorem}

\subsection{Generalized solutions}
Here we discuss possible generalizations of the exact solutions obtained above. 
As for the DWW equation, the idea is to replace the exponent of the exponentials (\ref{eq:phn}) and (\ref{eq:psn2}) by the infinite series 
\begin{equation}\label{infseries2}(x+3\k t^2)\la+bt\la^2+\sum_{j=3}^\infty \xi_j\la^j,\end{equation}
with $b=-1$ and $b=3$, respectively, where $\xi_j$, for $j\geq3$, are arbitrary parameters. However, when we substitute the generalizations of (\ref{eq:mbqsols123}) and (\ref{eq:mbqsols456}) into the \mbq\ equation (\ref{eq:mbq}), it can be shown that necessarily $\k=0$ and $\xi_j=0$ for $j\geq3$ except for the generalization of (\ref{eq:mbqsols456}) with $\k=0$, the derivatives of which decay as $|x|\to\infty$ and are decaying rational solutions of (\ref{eq:mbqnl}), and are described in the following theorem.

\begin{theorem}{Consider the polynomials $\th_n(x,t;\bfk)$ defined by
\begin{equation}\sum_{n=0}^\infty
\frac{\th_n(x,t;\bfk)\la^n}{n!}=\exp\left(x\la+3t\la^2+\sum_{j=3}^\infty
\xi_j\la^j\right),\end{equation} where $\bfk=(\xi_3,\xi_4,\ldots)$,
with $\xi_j$ arbitrary constants and then let 
\begin{equation}\label{eq:Psimnk}\begin{split}
\Theta_{m,n}(x,t;\bfk) &= %a_{m,n}
\mathcal{W}_x(\th_1,\th_{4},\ldots,\th_{3m+3n-5},
\th_2,\th_{5},\ldots,\th_{3n-1}),\\
\Theta_{-m,-n}(x,t;\bfk) &= %a_{m,n}
\mathcal{W}_x(\th_1,\th_{4},\ldots,\th_{3n-2},
\th_2,\th_{5},\ldots,\th_{3m+3n-1}),\end{split}\end{equation}
for $m,n\geq1$, where $\mathcal{W}_x(\theta_1,\theta_{2},\ldots,
\theta_{m})$ is the Wronskian with respect to $x$.
Then \mbq\ equation (\ref{eq:mbq}) has exact solutions in the form
\begin{subequations}\begin{align}\label{sols:mbqgenrats}
U_{m,n}^{[1]}(x,t;\bfk)&=\ln
\left\{\ifrac{\Theta_{m+1,n}(x,t;\bfk)}{\Theta_{m,n} (x,t;\bfk)}\right\},\\
U_{m,n}^{[2]}(x,t;\bfk)&=\ln
\left\{\ifrac{\Theta_{m,n}(x,t;\bfk)}{\Theta_{m,n+1} (x,t;\bfk)}\right\},\\
U_{m,n}^{[3]}(x,t;\bfk)&=\ln
\left\{\ifrac{\Theta_{m,n+1}(x,t;\bfk)}{\Theta_{m+1,n} (x,t;\bfk)}\right\}.
\end{align}\end{subequations} }\end{theorem}

\section{Discussion}\label{sec:Dis}
In this paper we have studied exact solutions of the DWW equation (\ref{eq:dww}) and the \mbq\ equation (\ref{eq:mbq}) through rational solutions of \PIV\ (\ref{eq:PIV}), which arises as a scaling reduction of these equations. Further we have derived some generalized solutions of the DWW equation (\ref{eq:dww}) and the \mbq\ equation (\ref{eq:mbq}) which are analogues of the generalized rational solutions of the KdV equation (\ref{eq:kdv}), the \bq\ equation (\ref{eq:bq}) 
and the NLS equation (\ref{eq:nls}). The DWW equation (\ref{eq:dww}) and the \mbq\ equation (\ref{eq:mbq}) also possess the accelerating wave reductions 
\begin{align}
U(x,t)&=W(z)+\mu xt-\tfrac23\mu^2t^3, \qquad z=x-\mu t^2,\label{dww:accwave}\\
U(x,t)&=W(z)-\tfrac13\mu xt, \qquad z=x-\mu t^2,\label{mbq:accwave}
\end{align}
respectively, where $\mu$ is an arbitrary constant and $W(z)$ is solvable in terms of solutions of \PII\ (\ref{eq:PII}). Since \PII\ has rational solutions expressed in terms of the \VY\ \cite{refVor,refYab59}, then we can derive another class of exact solutions of (\ref{eq:dww}) and (\ref{eq:mbq}) in terms of these special polynomials. We remark that the reduction (\ref{dww:accwave}) is obtained using classical Lie group method \cite{refBK,refOlver} whereas the reduction (\ref{mbq:accwave}) is a nonclassical symmetry reduction that is derived either using the nonclassical method \cite{refBCi} or the direct method \cite{refCK} --- see \cite{refPAC89mbq,refCL94} for details.

However, there are further exact solutions of the DWW equation (\ref{eq:dww}) and the \mbq\ equation (\ref{eq:mbq}), which are not special cases of the solutions discussed in \S\S3 and 4 above, or through the accelerating wave reductions (\ref{dww:accwave}) and (\ref{mbq:accwave}). For example, the DWW equation (\ref{eq:dww}) has the exact solutions
\begin{subequations}
\begin{align} u_1(x,t)&=\ln\left(\frac{x+t}{x+t+2}\right)-\frac{t}{2},\\
u_2(x,t)&=\ln\left\{\frac{(x+t)^3-3(x+t)^2-6t}{(x+t)^3+3(x+t)^2-6t}\right\}-\frac{t}{2},
\end{align}\end{subequations}
and the \mbq\ equation (\ref{eq:mbq}) has the exact solutions
\begin{subequations}
\begin{align} u_1(x,t)&=\ln\left(\frac{x+t}{x+t-6}\right)-\frac{t}{6},\\
u_2(x,t)&=\ln\left\{\frac{(x+t)^3+9(x+t)^2+24x+42t-144}{(x+t)^3-9(x+t)^2+24x+42t+144}\right\}-\frac{t}{6},
\end{align}\end{subequations}
which are not obtained by the procedure described above. These are analogous to the rational solutions of the classical Boussinesq equation (\ref{eq:clbq}) derived by
Sachs \cite{refSachs88} by applying the limiting procedure in \cite{refASat} to $N$-soliton solutions of (\ref{eq:clbq}).

The classical orthogonal polynomials, such as Hermite, Laguerre, Legendre,
and Tchebychev polynomials which are associated with rational solutions of the
classical special functions, play an important role in a variety of
applications (cf.\ \cite{refAbSt}). Hence it seems
likely that the special polynomials discussed here which are associated
with rational solutions of nonlinear special functions, i.e.\ the soliton
and \peqs, also arise in a variety of applications such as in numerical
analysis.

\section*{Acknowledgements} 
PAC thanks the organizers of the workshop on ``\emph{Group Analysis of Differential Equations 
and Integrable Systems}" held in Cyprus in October 2008 for the opportunity to give an invited lecture. BWMT acknowledges the support of an EPSRC studentship.
%I also thank Mark Ablowitz, Carl Bender, Bernard Deconinck, Galina Filipuk, Rod Halburd, Andrew Hone, Elizabeth Mansfield and Marta Mazzocco for their helpful comments and illuminating discussions.

\def\AAM{Acta Appl. Math.}
\def\ARMA{Arch. Rat. Mech. Anal.}
\def\bull{Acad. Roy. Belg. Bull. Cl. Sc. (5)}
\def\AC{Acta Crystrallogr.}
\def\AM{Acta Metall.}
\def\ampa{Ann. Mat. Pura Appl. (IV)}
\def\AP{Ann. Phys., Lpz.}
\def\APNY{Ann. Phys., NY}
\def\APP{Ann. Phys., Paris}
\def\BAMS{Bull. Amer. Math.Soc.}
\def\CJP{Can. J. Phys.}
\def\cmp{Commun. Math. Phys.}
\def\CMP{Commun. Math. Phys.}
\def\cpam{Commun. Pure Appl. Math.}
\def\CPAM{Commun. Pure Appl. Math.}
\def\CQG{Classical Quantum Grav.}
\def\crp{C.R. Acad. Sc. Paris}
\def\CSF{Chaos, Solitons \&\ Fractals}
\def\DE{Diff. Eqns.}
\def\DU{Diff. Urav.}
\def\ejam{Europ. J. Appl. Math.}
\def\EJAM{Europ. J. Appl. Math.}
\def\funk{Funkcial. Ekvac.}
\def\FUNK{Funkcial. Ekvac.}
\def\IP{Inverse Problems}
\def\JAMS{J. Amer. Math. Soc.}
\def\JAP{J. Appl. Phys.}
\def\JCP{J. Chem. Phys.}
\def\JDE{J. Diff. Eqns.}
\def\JFM{J. Fluid Mech.}
\def\JJAP{Japan J. Appl. Phys.}
\def\JP{J. Physique}
\def\JPhCh{J. Phys. Chem.}
\def\JMAA{J. Math. Anal. Appl.}
\def\JMMM{J. Magn. Magn. Mater.}
\def\JMP{J. Math. Phys.}
\def\jmp{J. Math. Phys.}
\def\JNMP{J. Nonl. Math. Phys.}
\def\jpa{J. Phys. A: Math. Gen.}
\def\JPA{J. Phys. A: Math. Gen.}
\def\JPB{J. Phys. B: At. Mol. Phys.} %1968-87
\def\jpb{J. Phys. B: At. Mol. Opt. Phys.} %1988 and onwards
\def\JPC{J. Phys. C: Solid State Phys.} %1968--1988
\def\JPCM{J. Phys: Condensed Matter} %1989 and onwards
\def\JPD{J. Phys. D: Appl. Phys.}
\def\JPE{J. Phys. E: Sci. Instrum.}
\def\JPF{J. Phys. F: Metal Phys.}
\def\JPG{J. Phys. G: Nucl. Phys.} %1975--1988
\def\jpg{J. Phys. G: Nucl. Part. Phys.} %1989 and onwards
\def\JSP{J. Stat. Phys.}
\def\JOSA{J. Opt. Soc. Am.}
\def\JPSJ{J. Phys. Soc. Japan}
\def\JQSRT{J. Quant. Spectrosc. Radiat. Transfer}
\def\LMP{Lett. Math. Phys.}
\def\LNC{Lett. Nuovo Cim.}
\def\NC{Nuovo Cim.}
\def\NIM{Nucl. Instrum. Methods}
\def\NL{Nonlinearity}
\def\NMJ{Nagoya Math. J.}
\def\NP{Nucl. Phys.}
\def\pl{Phys. Lett.}
\def\PL{Phys. Lett.}
\def\PMB{Phys. Med. Biol.}
\def\PR{Phys. Rev.}
\def\PRL{Phys. Rev. Lett.}
\def\PRS{Proc. R. Soc.}
\def\prsl{Proc. R. Soc. Lond. A}
\def\PRSL{Proc. R. Soc. Lond. A}
\def\PS{Phys. Scr.}
\def\PSS{Phys. Status Solidi}
\def\PTRS{Phil. Trans. R. Soc.}
\def\RMP{Rev. Mod. Phys.}
\def\RPP{Rep. Prog. Phys.}
\def\RSI{Rev. Sci. Instrum.}
\def\SAM{Stud. Appl. Math.}
\def\sam{Stud. Appl. Math.}
\def\SSC{Solid State Commun.}
\def\SST{Semicond. Sci. Technol.}
\def\SUST{Supercond. Sci. Technol.}
\def\TMP{Theo. Math. Phys.}
\def\ZP{Z. Phys.}
\def\OUP{O.U.P.} %{Oxford University Press}
\def\CUP{C.U.P.} %{Cambridge University Press}

\def\refpp#1#2#3#4#5{%\vspace{-0.25cm}
\bibitem{#1} \textrm{\frenchspacing#2}, \textrm{#3}, #4 (#5).}

\def\refjl#1#2#3#4#5#6#7{\vspace{-0.25cm}
\bibitem{#1}{\frenchspacing\rm#2}, %{\rm#6}, 
\textsl{\frenchspacing#3},  \textbf{#4} (#7) #5.}

\def\refjltoappear#1#2#3#4#5#6#7{\vspace{-0.25cm}
\bibitem{#1}{\frenchspacing\rm#2}, {\rm#6}, 
\textsl{\frenchspacing#3}, to appear.}

\def\refbk#1#2#3#4#5{\vspace{-0.25cm}
\bibitem{#1}{\frenchspacing\rm#2}, ``\textit{#3}", #4, #5.} 

\def\refcf#1#2#3#4#5#6{\vspace{-0.25cm}
\bibitem{#1} \textrm{\frenchspacing#2}, \textrm{#3},
in ``\textit{#4}"\ ({\frenchspacing#5}),  #6.}
\def\fit{\frenchspacing\it}

\end{document}